\documentclass[fleqn,usenatbib]{mnras}

\usepackage{newtxtext,newtxmath}
\usepackage[T1]{fontenc}

\DeclareRobustCommand{\VAN}[3]{#2}
\let\VANthebibliography\thebibliography
\def\thebibliography{\DeclareRobustCommand{\VAN}[3]{##3}\VANthebibliography}

\usepackage{graphicx}	% Including figure files
\usepackage{amsmath}	% Advanced maths commands
\usepackage{dirtytalk}

\title[Dust in scallop-shell stars]{Can scallop-shell stars trap dust in their magnetic fields?}

\author[H. Sanderson et al.]{
H. Sanderson,$^{1}$\thanks{E-mail: hannah.sanderson@earth.ox.ac.uk}
M. Jardine,$^{2}$
A. Collier Cameron,$^{2}$
J. Morin,$^{3}$
J.-F. Donati,$^{4}$\\
$^{1}$Department of Earth Sciences, University of Oxford, South Parks Road, Oxford, OX1 3AN
\\
$^{2}$SUPA, School of Physics and Astronomy, North Haugh, St Andrews, Fife KY16 9SS, UK\\
$^{3}$LUPM, Universit\'e de Montpellier, CNRS, Place Eug\`ene Bataillon, F-34095 Montpellier, France\\
$^{4}$Institut de Recherche en Astrophysique et Plan\'etologie, Universit\'e de Toulouse, UPS-OMP, 31400 Toulouse, France}

\date{Accepted 2022 November 08. Received 2022 October 11; in original form 2022 August 18.}

\pubyear{2022}

\begin{document}
\label{firstpage}
\pagerange{\pageref{firstpage}--\pageref{lastpage}}
\maketitle

\begin{abstract}
One of the puzzles to have emerged from the Kepler and TESS missions is the existence of unexplained dips in the lightcurves of a small fraction of rapidly-rotating M dwarfs in young open clusters and star-forming regions. We present a theoretical investigation of one possible explanation - that these are caused by dust clouds trapped in the stellar magnetic fields. The depth and duration of the observed dips allow us to estimate directly the linear extent of the dust clouds and their distances from the rotation axis. The dips are found to be between 0.4-4.8\%. We find that their distance is close to the co-rotation radius: the typical location for stable points where charged particles can be trapped in a stellar magnetosphere. We estimate the charge acquired by a dust particle due to collisions with the coronal gas and hence determine the maximum grain size that can be magnetically supported, the stopping distance due to gas drag and the timescale on which dust particles can diffuse out of a stable point. Using the observationally-derived magnetic field of the active M dwarf V374 Peg, we model the distribution of these dust clouds and produce synthetic light curves. We find that for 1$\mu$m dust grains, the light curves have dips of 1\% - 3\% and can support masses of order of $10^{12}$\,kg.  We conclude that magnetically-trapped dust clouds (potentially from residual disc accretion or tidally-disrupted planetesimal or cometary bodies) are capable of explaining the periodic dips in the Kepler and TESS data. 
\end{abstract}


%%%%%%%%%%%%%%%%%%%%%%%%%%%%%%%%%%%%%%%%%%%%%%%%%%

%%%%%%%%%%%%%%%%% BODY OF PAPER %%%%%%%%%%%%%%%%%%

\section{Introduction} \label{introduction}

Over the last few years, observations from K2 \citep{rebull_rotation_2016,stauffer_orbiting_2017,stauffer_more_2018,stauffer_even_2021} and TESS \citep{zhan_complex_2019} have revealed a class of rapidly-rotating ($P< 1$ d) M dwarfs with unusual asymmetric, rigidly-periodic light curves. The light curves have broad, quasi-sinusoidal modulations in amplitude, similar to those expected for starspots but with sharp smaller features superimposed, whose variability is too rapid to be explained by starspots alone \citep{stauffer_orbiting_2017,zhan_complex_2019}. The large and broad features in the light curves are stable for years whilst the smaller and sharper features can survive for a few months and disappear on a timescale of days \citep{zhan_complex_2019,gunther_complex_2022}. Over the course of a full year, several distinct manifestations of the scallop phenomenon appeared, each lasting between 1 and 2 months. The similarities (including strict periodicity and complex Fourier spectra) between the light curves of the scallop-shells discovered by \citet{stauffer_orbiting_2017} and the complex rotators discovered by \citet{zhan_complex_2019} suggest they are one class of star \citep{gunther_complex_2022}. They will be referred to as scallop-shells for the rest of this paper. These stars differ from the other classes of variable M dwarf, dippers and spotted stars with differential rotation, because they have shorter rotational periods and the variability is more rigidly periodic \citep{gunther_complex_2022}. 

The main theories suggested to explain these light curves are: co-rotating clouds of gas/dust at the Keplerian co-rotation radius ($R_K=\left(GM_\star / \Omega^2\right)^{1/3}$) superposed on an underlying rotational starspot modulation \citep{stauffer_orbiting_2017,stauffer_more_2018, gunther_complex_2022}, starspots with a misaligned dust disc \citep{zhan_complex_2019}, and eclipses of bound-free emission from gaseous slingshot prominences as they pass {\em behind} the star \citep{rebull_rotation_2016}.
%and spots with co-rotating clouds \citep{gunther_complex_2022}. 
The lack of an IR excess in the SEDs of these stars rules out a primordial, accreting dust disc as an explanation \citep{stauffer_orbiting_2017}. 

A recent study of occurrence rates among the complex rotational variables implies that, in order to explain these dips, some 30 percent of young rapidly-rotating M dwarfs would need to have inner dust discs misaligned with their rotation axes \citep{gunther_complex_2022}. We also note that for spot occultation by a dust disk to produce the observed sharp decreases in brightness, the occulted spots would have to be bright. Doppler-imaging studies of non-accreting M dwarfs \citep[e.g.][]{barnes_spots_2001,barnes_spots_2015}, however, invariably show significant coverage of dark rather than bright spots. This argues against models involving starspots and a misaligned disc.

Previous authors have discounted the co-rotating-clouds hypothesis, because gas clouds alone could not give the required absorption depth and it was claimed that the magnetic field at the co-rotation radius is too weak to trap dust \citep{gunther_complex_2022}.  However, studies of rapidly-rotating stars show that local traps exist within their magnetic fields where clouds of cool, dense gas (known as ``slingshot prominences'') may be supported \citep{2000MNRAS.316..647F,2001MNRAS.324..201J,2020MNRAS.491.4076J}. The existence of these mechanically-stable equilibrium points within stellar magnetic fields merits a closer look at the confinement of dust in similar locations.

Dust clouds may not be the only source of photometric variation. Spots may also contribute. Starspot activity is ubiquitous in such young, active stars; indeed the quasi-sinusoidal modulation of their light curves is the very phenomenon that allows their rotation periods to be measured photometrically. In the subset of these stars that exhibit scallop-shell behaviour, the short-duration absorption dips  repeat on the same period as the underlying starspot modulation. This indicates that the structures causing the dips co-rotate with the photosphere. Therefore, we propose the scallop-shell behaviour is caused by a superposition of starspots (the large quasi-sinusoidal modulation) and co-rotating clouds of dust trapped in magnetic stable points (the narrow features).  Magnetospheric gas \citep{gunther_complex_2022} or dust \citep{palumbo_evidence_2022} clouds have been briefly explored as an explanation but quantitative analysis of the dust trapping mechanism and the resulting light curves has not been done. 

We suggest these co-rotating dusty clouds are produced when dust (potentially from a tidally disrupted body analogous to a sun-grazing comet) becomes collisionally charged as it enters the corona, and couples to the magnetic field. The dust moves along magnetic field lines to magnetic stable points, whilst being decelerated by gas drag from the corona. Once there, the diffusion and sublimation timescales of dust at the stable points are slow, so the sharp features produced by the dust are stable over the observed timescale. Evolution of the stellar magnetic field (due to for example surface differential rotation or flux emergence) or the onset of thermal collapse leading to upflow of material into the stable point could lead to the eventual ejection of the dust cloud and disappearance of the narrow feature (see Section \ref{discussion:disappearance}).  

This paper provides a theoretical basis for this model. We consider the effect on the dust particles of both the magnetic field and plasma in the solar corona and demonstrate that these particles can be confined by the magnetic field in stable points. We use observational data from the complex rotators presented in \citet{zhan_complex_2019} to confirm the position of the dust clouds close to the co-rotation radius and determine the cloud sizes. Using these observational values we derive possible dust-cloud masses and use a magnetic map of an illustrative rapidly-rotating M dwarf (V374 Peg) to generate theoretical light curves caused by the subset of these magnetospheric clouds that transit the stellar disc. 

The structure of the paper is as follows: Section \ref{theory} outlines the methods for extracting information from the observations and the theory for the confinement of dust at magnetic stable points. Section \ref{results} applies the methods and theory from Section \ref{theory} to observed scallop-shell stars and V374 Peg and generates theoretical light curves for magnetospherically trapped dust. Section \ref{discussion} discusses the implications of these results and we conclude in Section \ref{conclusion}.

\section{Theory and Methods}\label{theory}

\subsection{Observational indicators of dust-cloud locations and sizes} 
\label{methods:observations}

These \citet{stauffer_orbiting_2017,stauffer_more_2018} M dwarfs all have spectral types M5 or later, and rotation periods of order half a day or less. Given that the transients recur on the same period as the starspot modulation, they must co-rotate with the photosphere. This places them either in Keplerian orbits synchronous with the stellar rotation, or confined in co-rotating structures.
Given that the transients arise when co-rotating structures transit the stellar disc, their durations depend on their distances from the stellar rotation axis. The dip duration is the length of time taken for an isolated cloud to traverse the stellar diameter at the transverse rotational speed of its location. Expressing the dip duration $T$ as a fraction of the rotation period $P$,
\begin{equation}
    \frac{T}{P}\simeq\frac{2R_\star}{2\pi a_{\rm cloud}},
\end{equation}
where $a_{\rm cloud}$ is the distance of the cloud from the stellar rotation axis. This assumes that the cloud transits the centre of the disc. Shorter durations may be seen if the path of the cloud passes well away from the disc centre.

\citet{zhan_complex_2019} identified 10 scallop stars in their TESS survey from the existence of high-frequency harmonics of the rotation period in the Fourier transforms of their light curves. If the highest frequency detectable above the noise is related to the shortest transient duration by $\nu_{\rm max}\simeq1/T_{\rm min}$, we can establish a lower limit on $a_{\rm cloud}$ either from the Fourier transform of the light curve or by measuring the duration of the shortest distinct dip seen in the light curve. Given an estimate of the stellar mass and rotation period, we can then compare the inferred value of  $a_{\rm cloud}$ directly with the Keplerian co-rotation radius. For the results given in Section \ref{results:observations} we used the same sample of 10 TESS light curves examined by \citet{zhan_complex_2019} and by \citet{gunther_complex_2022}.

The cloud radius $r_{\rm cloud}$ can be estimated by considering the  observed fractional depth $k$ of an isolated cloud transient (typically $k=2-4\%$). Empirically, this tells us the ratio \begin{equation}\label{eq:rcloud}
        k=\frac{A_{\rm cloud}}{A_\star} = \left(\frac{r_{\rm cloud}}{R_\star}\right)^2.
\end{equation} Hence $r_{\rm cloud}=\sqrt{k}R_\star$ where $k$ is the fractional flux deficit at mid-event for a small cloud.

\subsection{Magnetic field structure and the position of dust clouds}\label{methods:magnetic_field_structure} %Moira 
%Aim: explain concept of stable point and explain how they are modelled
One of the main challenges in understanding this phenomenon is to explain how these clouds can be confined in discrete locations and forced to co-rotate with the star. This is possible within a magnetic field if the cloud is located close to a stable point in the field. These are local gravitational potential minima {\it as calculated following the direction of the field}. These locations can be determined if the magnetic field structure is known \citep{2000MNRAS.316..647F,2001MNRAS.324..201J}. They are defined by
\begin{equation}
\left( \underline{B} \cdot \underline{\nabla} \right) \left(  \underline{g}_{\rm eff} \cdot \underline{B} \right) < 0.
\end{equation}
Here the effective gravitational acceleration is given by $\underline{g}_{\rm eff} =( \underline{g}.\underline{B})/|\underline{B}|$ and 
\begin{equation}
\underline{g}(r,\theta) = \left( -GM_{\star}/r^{2} + 
                     \omega_\star^{2}r\sin^{2}\theta,
		     \omega_\star^{2}r\sin\theta\cos\theta 
             \right), 
\end{equation}
in spherical coordinates (r, $\theta$, $\phi$), with the polar direction aligned with the rotation axis of the star, where $\omega_\star$ is the stellar angular velocity. For example, for a simple dipolar magnetic field where the dipole axis is aligned with the rotation axis, the stable points lie in the rotational equator at radii $r > (2/3)^{1/3}R_K$ \citep{2000MNRAS.316..647F}. Even for more complex field structures, the stable points typically cluster close to, or just beyond, the co-rotation radius \citep{2001MNRAS.324..201J}.

In order to determine the coronal structure of the magnetic field, we can extrapolate from maps of the surface magnetic field obtained using Zeeman-Doppler imaging \citep{1997MNRAS.291....1D,1999MNRAS.302..437D,2000MNRAS.318..961H,2012A&A...548A..95C,2015ApJ...805..169R}. Here we assume that the coronal field is potential and therefore supports zero volume currents. We can then use the well-established {\it Potential Field Source Surface} method \citep{1969SoPh....9..131A} to perform the extrapolation. Since the field is potential we may express it as $\underline{B}=-\underline{\nabla} \Psi$, with the result that $\underline{\nabla} \cdot \underline{B} = 0$ then requires $\nabla^2 \Psi = 0$. We may therefore express $\Psi$ 
in spherical co-ordinates $(r,\theta,\phi)$ as
\begin{equation}
 \Psi = \sum_{l=1}^{N}\sum_{m=-l}^{l} [a_{lm}r^l + b_{lm}r^{-(l+1)}]
         P_{lm}(\theta) e^{i m \phi},
\label{eqn:Laplace}
\end{equation}
where all radii are scaled to the stellar radius. The map of the surface radial magnetic field therefore provides one of the boundary conditions for Eqn. \ref{eqn:Laplace}. For the other, we assume that at some radius (know as the source surface, r$_{ss}$) the field is is forced to become purely radial by the pressure of the hot coronal gas, and so at $r=r_{ss}$, $B_\theta=B_\Phi = 0$. Hence 
\begin{equation}
b_{lm}=-a_{lm} r_{ss}^{2l+1}.
\end{equation}
For further details, the reader is referred to the Appendix.

\begin{figure}
    \centering
    \includegraphics[width=1\columnwidth]{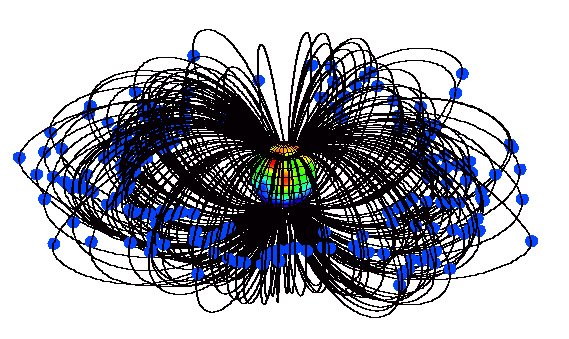}\\
    \includegraphics[width=1\columnwidth]{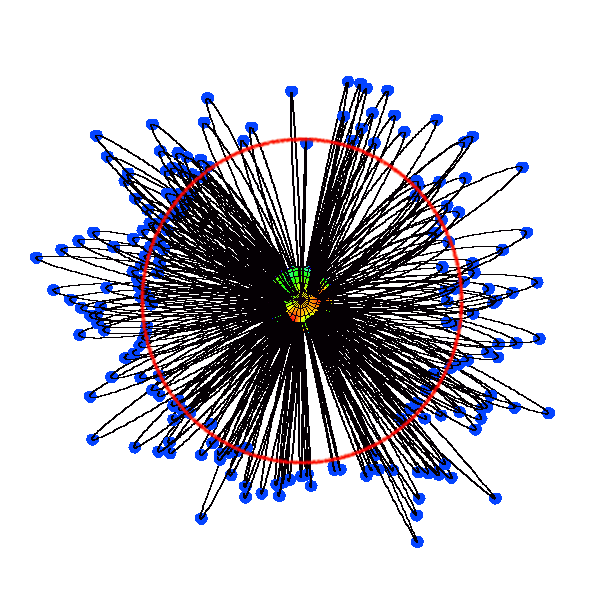}
    \caption{Magnetic field structure of V374 Peg with dust-cloud locations shown as blue clumps. The top panel shows the view as see from the observer's viewpoint at inclination 70$^o$ (top) while the bottom panel (where the co-rotation radius is marked as a red circle) shows the view from above the rotation axis. The radial magnetic field map is shown on the stellar surface (blue is negative, red is positive polarity).}
    \label{fig:stable_points}
\end{figure}

We show an example of such a field structure and the associated stable points in Fig.~\ref{fig:stable_points}. This field extrapolation utilises the surface magnetic field map of the rapidly-rotating M dwarf V374 Peg \citep{2006Sci...311..633D}. The stable points cluster around the co-rotation radius, as would be expected.

\subsection{Dust particle confinement in the magnetic field} \label{methods:trapping_model}
Magnetospherically-trapped dust has previously been suggested by \citet{farihi_magnetospherically-trapped_2017} to explain the unusual pattern of evolving, short-duration periodic transits observed by \citet{vanderburg_disintegrating_2015} and \citet{gansicke_high-speed_2016} in the polluted white dwarf
WD1145+017. However, a model for trapping charged dust around M dwarfs differs from that for white dwarfs, because M dwarfs are larger and have an extensive corona. Collisions with ions in the corona exert drag on the dust particles and are the main cause of dust-particle ionisation due to their high number density ($10^{16}-10^{18}\rm m^{-3}$\citet{ness_coronal_2002,ness_sizes_2004}), in contrast to white dwarfs where photoionisation dominates.

We propose that the confinement of dust in the stable points occurs as follows. 
The dust grains initially move towards the star from an external source (see Section \ref{discussion}) due to a combination of gravity and Poynting-Robertson drag. As the dust particle enters the stellar corona it will rapidly become charged by collisions with the plasma and therefore will experience the Lorentz force in addition to drag from the coronal plasma. The density of this plasma ensures that once inside the corona, the radiation force experienced by the dust particles is negligible in comparison to the collisional drag on the dust grains. If the Lorentz force exceeds the gravitational force, the magnetic field may be able to capture the dust grains. Even in this case, however, dust particles will be free to move along field lines. If they encounter a stable location they will remain there. The effect of collisional drag with the coronal gas will be to slow down the motion of the dust grain, potentially preventing it from reaching a stable point. Typically stable points are close to the co-rotation radius.
\subsubsection{Charging of dust grains}
In order for the dust particles to be affected by the magnetic field, they must be charged. The dust grains become charged by collisions with electrons and protons in the coronal plasma. The dust grains are assumed to acquire a negative charge, because electrons in the plasma have faster velocities than electrons due to their lower masses so collide more frequently with the dust grains. The resulting charge, $q_d$ and collision cross section can be calculated following a method of \citet{ke_rapid_2012} to give \begin{equation} Q=-\frac{30{\pi}{\epsilon}_0r_{\rm d}k_{\rm B}T}{e}, \end{equation} where $k_{\rm B}$ is the Boltzmann constant, $\epsilon_0$ is the permittivity of free space, $e$ is the charge on one electron, $r_{\rm d}$ is the radius of the spherical dust grain and T is the plasma temperature. 
\subsubsection{Lorentz force}
For a charged dust grain in a dipole field, the magnitude of the Lorentz force is \begin{equation} \begin{split} F_L&=q_dv_{\rm d}B \\&=\frac{30 \pi \epsilon_0 r_{\rm d} k_{\rm B} T}{e}v_{\rm d} B_{\rm surf} \left(\frac{R_\star}{r}\right)^3 \end{split}\label{eq:lorentz}\end{equation} where $q_d$ is the charge on the dust grain, $v_{\rm d}$ is its velocity perpendicular to the magnetic field, $B_{\rm surf}$ is the surface magnetic field strength of the star, $r$ is the distance from the centre of the star and $R_\star$ is the stellar radius. For a dust grain moving towards the star, its velocity will be a fraction of the escape speed $v_{\rm d}=fv_{\rm esc}=f\sqrt{\frac{2GM_\star}{r}}$ where $0\leq f\leq 1$ is a factor that accounts for the alignment of the particle's velocity relative to the magnetic field and its velocity relative to the escape velocity. For example, a particle in a circular orbit has $v_{\rm d}=\frac{v_{\rm esc}}{\sqrt{2}}$. Equation~\ref{eq:lorentz} can therefore be rewritten as\begin{equation}
    F_L=\frac{30 \pi \epsilon_0 r_{\rm d} k_{\rm B} T}{e} f\sqrt{2GM_\star}B_{\rm surf}R_\star^3r^{-\frac{7}{2}}.
\end{equation}  
\subsubsection{Gravitational force and maximum grain size}
The gravitational force on a dust grain is given by \begin{equation}
    F_G=\frac{4\pi r_{\rm d}^3 \rho_{\rm d}}{3}\frac{GM_\star}{r^2}
\label{eq:gravity}\end{equation}
where $M_\star$ is the stellar mass, $\rho_{\rm d}$ is the bulk density of the dust grains and $G$ is the gravitational constant.
For a dust particle to be supported by the magnetic field against gravity we require $F_L>F_G$. The gravitational force is used rather than the effective gravitational force, because it is a more conservative criterion which takes into account that the dust is not initially on a circular orbit.
In our model, the dust grains are treated as spherical and we follow \citet{zhan_complex_2019} in using a typical density of rocky material of 3000\,kg\,$\rm m^{-3}$. The gravitational force on a particle scales as $r_{\rm d}^3$ whereas the charge (and Lorentz force) scale as $r_{\rm d}$. Therefore grains must be below a critical radius to be significantly affected by the magnetic field. Combining Equations \ref{eq:lorentz} and \ref{eq:gravity} and assuming that the particle is located at a radial distance $r=\beta R_K$ gives 
\begin{equation}
    \begin{split}
        r_{\rm d}<\left(\frac{45\sqrt{2}}{2}\frac{\epsilon_0k_{\rm B}T R_\star^3fB_{\rm surf}\Omega}{eGM_\star\rho_{\rm d}\beta^{\frac{3}{2}}}\right)^{\frac{1}{2}}.
    \end{split}\label{eq:gravity_lorentz_ratio}
\end{equation}
\subsubsection{Drag from coronal plasma} 
The drag force experienced by the dust grain due to collisions with protons in the plasma is \begin{equation}
    F_{\rm drag}= m_{\rm p}n_{\rm p}\sigma_{\rm col} c_s v_{\rm rel}
\end{equation} where $c_s=\sqrt{\frac{k_{\rm B}T}{\mu m_{\rm p}}}$ is the sound speed in the plasma and $\mu=0.61$ is the mean molecular weight of the plasma. This was approximated by primordial solar abundances by mass fraction \citep{asplund_chemical_2009}, which are a reasonable approximation for these young convective stars which will be well mixed. $\sigma_{\rm col}$ is the collision cross section between dust and protons in the plasma and $v_{\rm rel}$ is the relative velocity between the dust and plasma - again taken to be $fv_{\rm esc}$.
The collision cross section between a charged dust grain and protons is enhanced due to their electrostatic interaction. Following \citet{ke_rapid_2012}, this enhancement can be written as $\sigma_{\rm col}=\pi r_{\rm d}^2(1+\theta)$ and $\theta=5$ was used. Combining these expressions
\begin{equation}
    F_{\rm drag}=6\pi r_{\rm d}^2n_{\rm p}\sqrt{\frac{2k_{\rm B}Tm_{\rm p}GM_\star}{\mu r}}f.
    \label{eq:proton_collisions}\end{equation}
This drag force has a comparable magnitude to the Lorentz force for micrometer size grains. The drag force decelerates the dust particle, decreasing the Lorentz force it experiences. The stopping time, $\tau_{\rm s}$ gives an estimate of how long the particle will be influenced by the magnetic field before it is brought to rest relative to the gas. The stopping time is given by \begin{equation}
    \begin{split}
        \tau_{\rm s}&=\frac{m_{\rm d} v_{\rm d}}{F_{\rm drag}}
        \\&=\frac{m_{\rm d}}{n_{\rm p}m_{\rm p}c_s\sigma_{\rm col}}
    \end{split}
\end{equation} where $m_{\rm d}=\frac{4}{3}\pi r_{\rm d}^3 \rho_{\rm d}$ is the mass of a spherical dust grain. In turn, a stopping distance for a dust grain initially travelling at some fraction of the escape velocity can be estimated as \begin{equation}\begin{split}d_{\rm s}&=\frac{v_{\rm d}\tau_{\rm s}}{2}\\&=\frac{ r_{\rm d} \rho_{\rm d}f}{9n_{\rm p}}\sqrt{\frac{2GM_\star\mu}{m_{\rm p}k_{\rm B}Tr}} \\&=\frac{fr_d\rho_d(GM_*\Omega)^{\frac{1}{3}}}{9n_p}\sqrt{\frac{2\mu}{m_pk_BT\beta}},\end{split}\label{eq:stopping_distance}\end{equation} where the third equality comes from substituting $r=\beta\,R_K$.
This stopping distance must be large enough that dust grains can travel along field lines to a stable point once they enter the corona. Larger grains have larger stopping distances, because although they experience more drag they have higher inertia. 
\subsubsection{Sublimation and diffusion timescales}\label{sub-time}
The features observed in the light curves are stable over an 81 day K2 campaign \citep{stauffer_orbiting_2017} and 1-2 sectors (27-54+ days) of TESS observations \citep{zhan_complex_2019}. Therefore, once at the stable point, dust grains must have diffusion and sublimation timescales longer than a few months to remain at the stable point, assuming there is no resupply of dust. 
For sublimation timescales of dust grains close to M dwarfs we refer to the recent paper by \citet{zhan_complex_2019}, who considered five different compositions and three grain sizes (0.1$\mu$m, 1$\mu$m, 10$\mu$m) at distances of 1-7$R_\star$. For V374 Peg the co-rotation radius is at $R_K=4.72R_\star$, where three (corundum, forsterite and enstatite) of the five compositions considered by \citet{zhan_complex_2019} for all sizes and 10$\mu$m iron grains have sublimation timescales greater than a year and one (fayalite) has a sublimation timescale of one month. These timescales are sufficiently long that dust grains can survive long enough for the features observed and sublimation should not be the dominant contributor to the variability. 

Random thermal motion of particles within the plasma may of course lead to diffusion of dust out of the stable point. The diffusion coefficient can be estimated as $D=\frac{1}{2}\lambda v_{\rm p}$ where $\lambda=\frac{1}{n_{\rm p}\sigma_{\rm col}}$ is the mean free path for collisions between dust grains and protons in the plasma (since protons have greater momentum than the electrons in the plasma) and $v_p=\sqrt{\frac{3k_BT}{m_p}}$ by equipartition theorem. Assuming that at the stable point the dust is diffusing out of a spherical cloud of radius $r_{\rm cloud}$ and surface area $A=4\pi r_{\rm cloud}^2$, the diffusion timescale, $\tau_{\rm diff} = A/D$, can be written as  \begin{equation}\begin{split}
    \tau_{\rm diff} = \frac{8\pi r_{\rm cloud}^2n_{\rm p}\sigma_{\rm col}}{v_{\rm p}}.
\end{split}\label{eq:diffusion}\end{equation}

\subsection{Dust masses for the absorption and light curves} 
\label{subsection:dust_masses}
%Hannah and Moira
%Aim: explain how mass of dust required for absorption is calculated, explain how light curves are generated
The mass of dust required to produce the observed absorption must be physically realistic. To produce the observed absorption, the optical depth, $\tau$ of the dust cloud must be at least one. The optical depth is given by \begin{equation}
    \tau=n_{\rm d}\sigma_{\rm ext} s
\end{equation} where $n_{\rm d}$ is the number density of the absorbing species, $\sigma_{\rm ext}$ is the extinction cross section and $s$ is the path length. For micrometer size grains in the Kepler passband (400-900\,nm), light undergoes Mie scattering and the extinction cross section can be written as $\sigma_{\rm ext}=Q_{\rm ext}\pi r_{\rm d}^2$ (assuming spherical dust grains). $Q_{\rm ext}$ is a numerical prefactor that depends on the complex refractive index of the dust grains and the dimensionless parameter $X=\frac{2\pi r_{\rm d}}{\lambda}$, which characterises the size of the particle compared to the wavelength of the light, $\lambda$. \citet{croll_multiwavelength_2014} calculated $Q_{\rm ext}$ as a function of $X$ for a variety of refractive indices, including those for the same minerals considered by \citet{zhan_complex_2019} which are likely to be in the dust clouds. For values of $X$ greater than five, Figure 13 in \citet{croll_multiwavelength_2014} indicates there is little variation in $Q_{\rm ext}$ with increasing $X$ and composition has little effect. Therefore, for the optical depth calculation, the light was taken to be monochromatic at 650nm as it did not affect the extinction cross section. The path length for the optical depth calculation was taken to be twice the radius of the spherical dust cloud, such that $s = 2 r_{\rm cloud}$. This maximises the path length for a given cloud. The number density of the dust was taken to be constant within the cloud, because the dust clouds are small in comparison to the pressure scale height of the corona. This allows us to write
\begin{equation}
    n_{\rm d}=\frac{\tau}{2Q_{\rm ext}\pi r_{\rm d}^2 r_{\rm cloud}}.
\label{eq:number_densities}\end{equation}
The {\it minimum} dust number density can be obtained by setting $\tau=1$. 

The dust-cloud locations must be estimated to calculate model light curves. As discussed in Section \ref{methods:trapping_model}, dust clouds can be located at stable points (see Section \ref{methods:magnetic_field_structure}) in the magnetic field. At any given time, not all stable points will support dust clouds, so we randomly select half of the stable points  to support dust clouds and calculate multiple realisations of the resulting dust-cloud locations. We show one example in Fig. \ref{fig:stable_points}.  Equations \ref{eq:rcloud} and \ref{eq:number_densities} can then be used to calculate the cloud radii and number densities and can be combined with these dust-cloud locations to generate model light curves (see Section \ref{sec:V374LC}).

\begin{table}
    \centering
    \begin{tabular}{|c|c|}
       Quantity  & Value \\ \hline
       $T$  & 5.7\,MK 
       \\$M_\star$ & $0.28 M_\odot$ 
       \\ $R_\star$ & $0.34 M_\odot$ 
       \\$R_K$ & $4.72R_\star$ 
       \\ $B_{\rm surf}$ & 500G 
       \\ $\Omega_*$ & $ 1.63\times10^{-4}\,\rm rad \,\rm s^{-1}$ 
       \\ $\rho_{\rm d}$ & 3000 kg$\, \rm m^{-3}$
       \\ $n_{\rm p}$ & $10^{16}\, \rm m^{-3}$ 
       \\ $r_{\rm cloud}$ & $0.1\,R_\star$
       \end{tabular}
    \caption{Typical values for quantities in scaling relationships in this paper. Stellar parameters except temperature are values for V374 Peg from \citet{2006Sci...311..633D}, with $R_\star$ assuming sin$i=70$ \textdegree. The coronal temperature of V374 Peg comes from \citet{jardine_slingshot_2019}. The dust particle density follows \citet{zhan_complex_2019} and the proton number density in the corona comes are order of magnitude estimates based on \citet{ness_coronal_2002,ness_sizes_2004}. The cloud radius was obtained from the amplitude of the scallop-shell features (see Section \ref{results:masses}).}
    \label{tab:typical_values}
\end{table}

\section{Results}\label{results}

\subsection{Dust-cloud locations and sizes from observations}\label{results:observations} 

\begin{figure}
    \centering
    \includegraphics[width=\columnwidth]{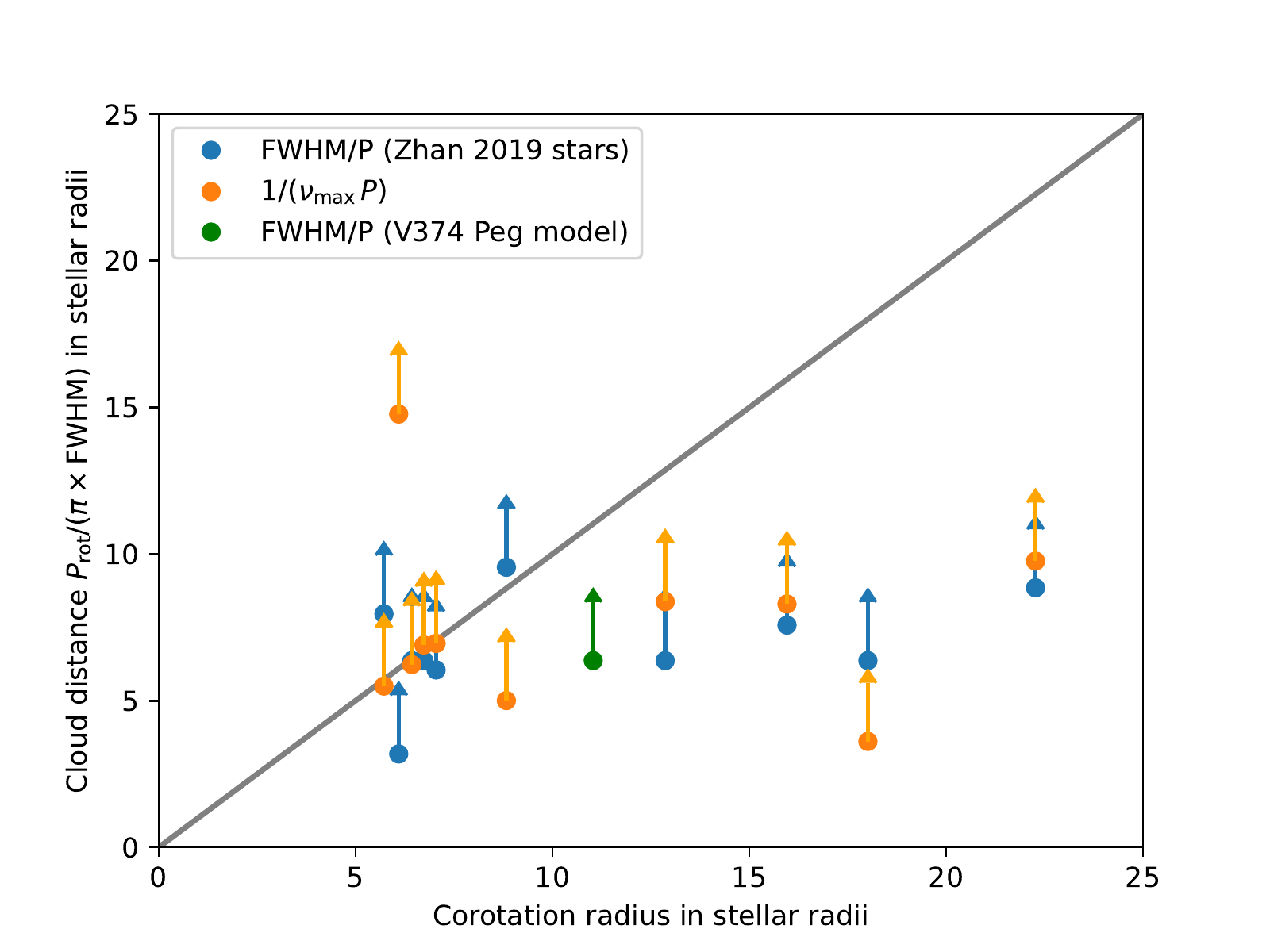}
    \caption{Lower limits on the radial distances $r/R_\star$ of the co-rotating clouds from the rotation axes of the 10 stars studied by \citep{zhan_complex_2019}, versus Keplerian co-rotation radius. The blue points are measured using the FWHM of the smallest dip feature in each light curve, while the orange points are derived from the highest frequency clearly visible in the Fourier power spectrum of the light curve. The grey line of unit slope allows the cloud locations to be compared with the co-rotation radius. 
    The green point, derived in the same way from the duration of the narrowest dip produced by clouds near the co-rotation radius in the synthetic light curve of V374 Peg (Fig.~\ref{fig:v374peg_light_curves}), illustrates the lower-limit nature of these measurements.}
    \label{figure:distances}
\end{figure}

In Table~\ref{table:cloudparams} we list the lower limits on the radial distances $a/R_\star$ of the co-rotating clouds from the rotation axes of the 10 stars studied by \citet{zhan_complex_2019} via the two different methods described in Section~\ref{methods:observations} and the radii of the dust clouds. The dips are between 0.4-4.8\% corresponding to cloud radii of 0.06-0.22 $R_\star$. We identified the narrowest dip in each of the published light curves and measured its fractional depth $k$ and full duration $T$ at half maximum depth directly. We also computed the Fourier power spectrum of the TESS data for each of the \citet{zhan_complex_2019} targets, and determined the timescale corresponding to the highest-frequency peak in the regularly-spaced pattern clearly visible above the noise. Both methods yielded similar estimates for $T$, and hence for $r/R_\star = P/\pi T$. The distances $r$ of the clouds from the rotation axis are plotted and compared with the Keplerian co-rotation radius in Fig.~\ref{figure:distances}.

With the exception of one outlier, which could be the product of an elongated cloud complex with an extent greater than the stellar diameter, the narrowest clearly-identifiable dips in all the stars in the sample of \citet{zhan_complex_2019} lie 
between 5 and 10 stellar radii from the rotation axis. These should be considered as lower limits, since the narrowest events may result from grazing transits of clouds in an equatorial ring tilted relative to the line of sight. As with transiting planets, the probability that a randomly-inclined object at distance $r$ will transit the stellar disc is $\mathcal{P}(\rm transit) = R_\star/r$. This produces a geometrical selection effect resulting from the random inclinations of the stellar rotation axes, favouring grazing transits when the co-rotation radius is further from the stellar rotation axis. We validated this explanation by using the dip widths in the synthetic light curve of V374 Peg discussed and illustrated in Section \ref{sec:V374LC}, to estimate the radial distance in the same way. The derived distance, shown as a green lower limit in Fig.~\ref{figure:distances}, was about 60 percent of the co-rotation radius, whereas the model clouds were all close to co-rotation. We conclude that the dips in these stars arise from co-rotating structures predominantly located close to, or inside, the co-rotation radius.

\begin{table*}
\caption{Estimated cloud sizes and radial distances for the smallest dip features identified in TESS light curves of the ten 10 stars from \citep{zhan_complex_2019}. The first 6 columns give the TIC number , stellar mass and radius in solar units, the rotation period in days, the fractional duration $T/P$ of the smallest dip, and its fractional depth $k$. The next three columns list the Keplerian co-rotation radius, minimum cloud distance $r_{\rm FWHM}$ estimated from the FWHM of the smallest dip, and the minimum cloud distance $r_{\rm Fourier}$ estimated from the highest frequency in the Fourier power spectrum of the light curve. The final column gives the cloud size $r_{\rm cloud}$ estimated from the fractional depth of the smallest dip.}
\label{table:cloudparams}

\begin{tabular}{lrrrrrrrrr}
\hline\\
      TIC &    $M_\star/M_\odot$ &    $R_\star/R_\odot$ &      $P_{\rm rot}$ (day) & $T/P$ & $k$ &     $R_{\rm K}/R_\star$ & $r_{\rm FWHM}/R_\star$ & $r_{\rm Fourier}/R_\star$ &                 $r_{\rm cloud}/R_\star$ \\
\hline\\
206544316 & 0.35 & 0.46 & 0.3217 & 0.100 & 0.048  &  6.1 &  3.2 & 14.8 & 0.22 \\
224283342 & 0.34 & 0.25 & 0.8873 & 0.036 & 0.019  & 22.3 &  8.9 &  9.8 & 0.14 \\
425933644 & 0.38 & 0.51 & 0.4863 & 0.053 & 0.036  &  7.0 &  6.0 &  7.0 & 0.19 \\
425937691 &  0.3 & 0.32 & 0.2007 & 0.050 & 0.010  &  6.7 &  6.4 &  6.9 & 0.10 \\
 38820496 & 0.25 & 0.28 &  0.656 & 0.050 & 0.008  & 18.0 &  6.4 &  3.6 & 0.09 \\
201789285 & 0.18 & 0.24 & 0.1516 & 0.033 & 0.018  &  8.8 &  9.6 &  5.0 & 0.13 \\
234295610 & 0.37 & 0.38 & 0.7615 & 0.050 & 0.005  & 12.9 &  6.4 &  8.4 & 0.07 \\
177309964 & 0.49 & 0.49 & 0.4533 & 0.050 & 0.014  &  6.4 &  6.4 &  6.3 & 0.12 \\
332517282 & 0.19 & 0.25 & 0.4023 & 0.042 & 0.010  & 16.0 &  7.6 &  8.3 & 0.10 \\
289840928 & 0.29 & 0.38 & 0.1999 & 0.040 & 0.003  &  5.7 &  8.0 &  5.5 & 0.06 \\
\hline\\
\end{tabular}

\end{table*}

\subsection{Can the dust be confined by the magnetic field?}\label{results:dust_confinement}

Typical values for dust and stellar parameters can be substituted into the equations from Section \ref{methods:trapping_model} to determine if dust trapping in magnetic stable points is possible. The typical values used are summarised in Table \ref{tab:typical_values} and stellar parameters have been taken for the magnetically active M dwarf V374 Peg, but the results here are presented as scaling relationships so they can be applied to other stars. Although V374 Peg does not show the lightcurve dips that are characteristic of the scallop-shell stars, it was used as an illustrative example of the magnetic field structures that we might expect to find on rapidly-rotating mid-to-late M dwarfs. It is well observed \citep{2006Sci...311..633D,2008MNRAS.384...77M,2009AIPC.1094..146H,vida_investigating_2016} and magnetic field maps are available for it, which enables the modelling of the stable points discussed in Section \ref{methods:magnetic_field_structure}. 

From Equation \ref{eq:gravity_lorentz_ratio} the critical size below which Lorentz forces dominate over gravitational ones is

 \begin{equation}
 r_{\rm d}<12\frac{\left(\frac{f}{1}\right)^{\frac{1}{2}}\left(\frac{T}{5.7\,\rm M \rm K}\right)^{\frac{1}{2}} \left(\frac{R_\star}{0.34R_\odot}\right)^{\frac{3}{2}}\left(\frac{B_{\rm surf}}{500\, \rm G}\right)^{\frac{1}{2}}\left(\frac{\Omega}{1.6\times10^{-4}\, \rm rad\,\rm s^{-1}}\right)^{\frac{1}{2}}}{\left(\frac{M_\star}{0.28M_\odot}\right)^{\frac{1}{2}}\left(\frac{\rho_{\rm d}}{3000\,\rm kg\, \rm m^{-3}}\right)^{\frac{1}{2}}\left(\frac{\beta}{1}\right)^{\frac{3}{4}}}\mu  \rm m
    \end{equation}

\begin{figure*}
    \centering
    \includegraphics[width=1\textwidth]{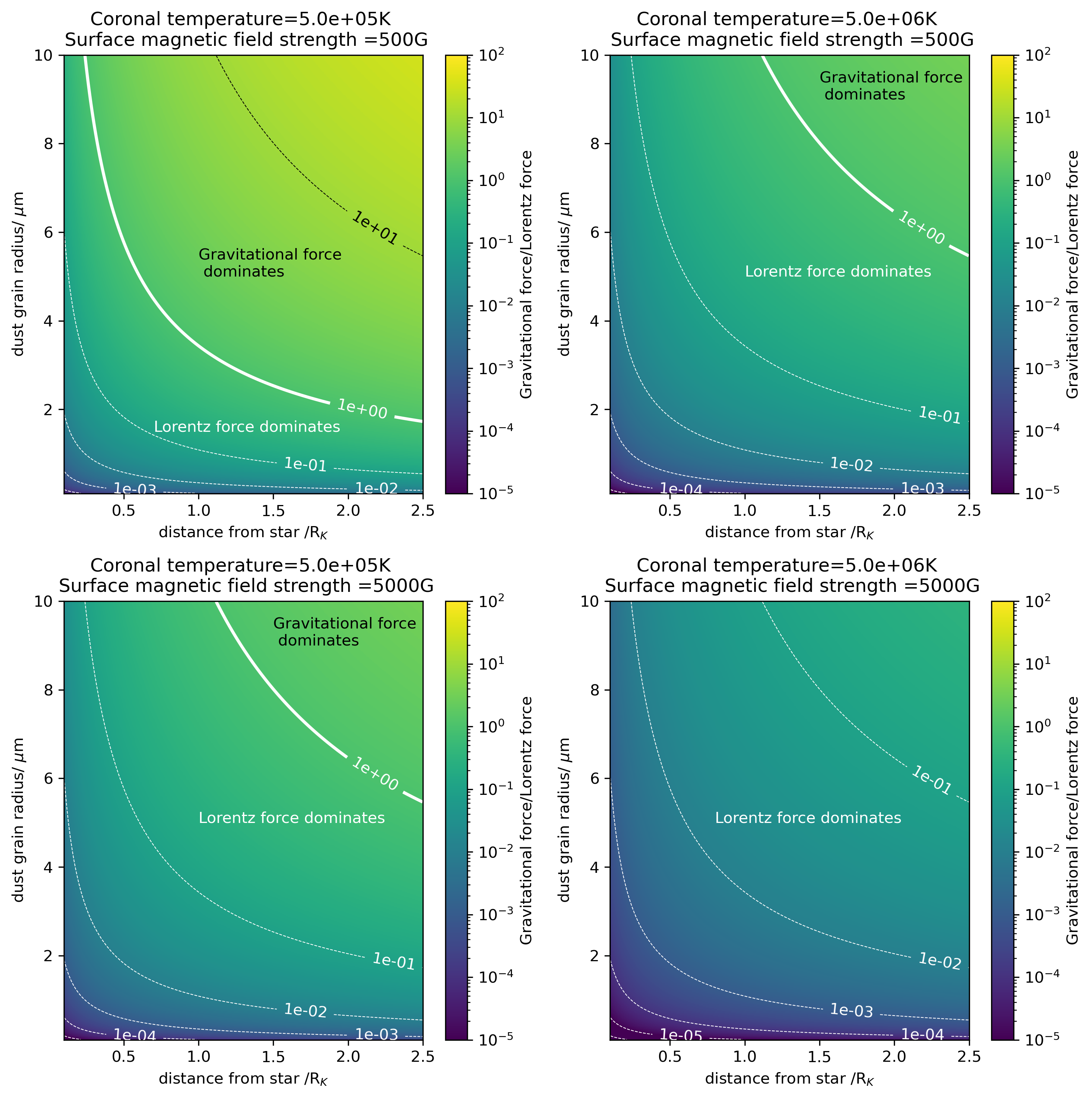}
    \caption{Dust grain support as a function of grain size and distance from the star, for a variety of coronal temperatures and surface field strengths. The temperature down each column of the subplot is constant and the surface field strength across each row of the subplot is constant.}
    \label{fig:support-ar}
\end{figure*}

\begin{figure*}
    \centering
    \includegraphics[width=1\textwidth]{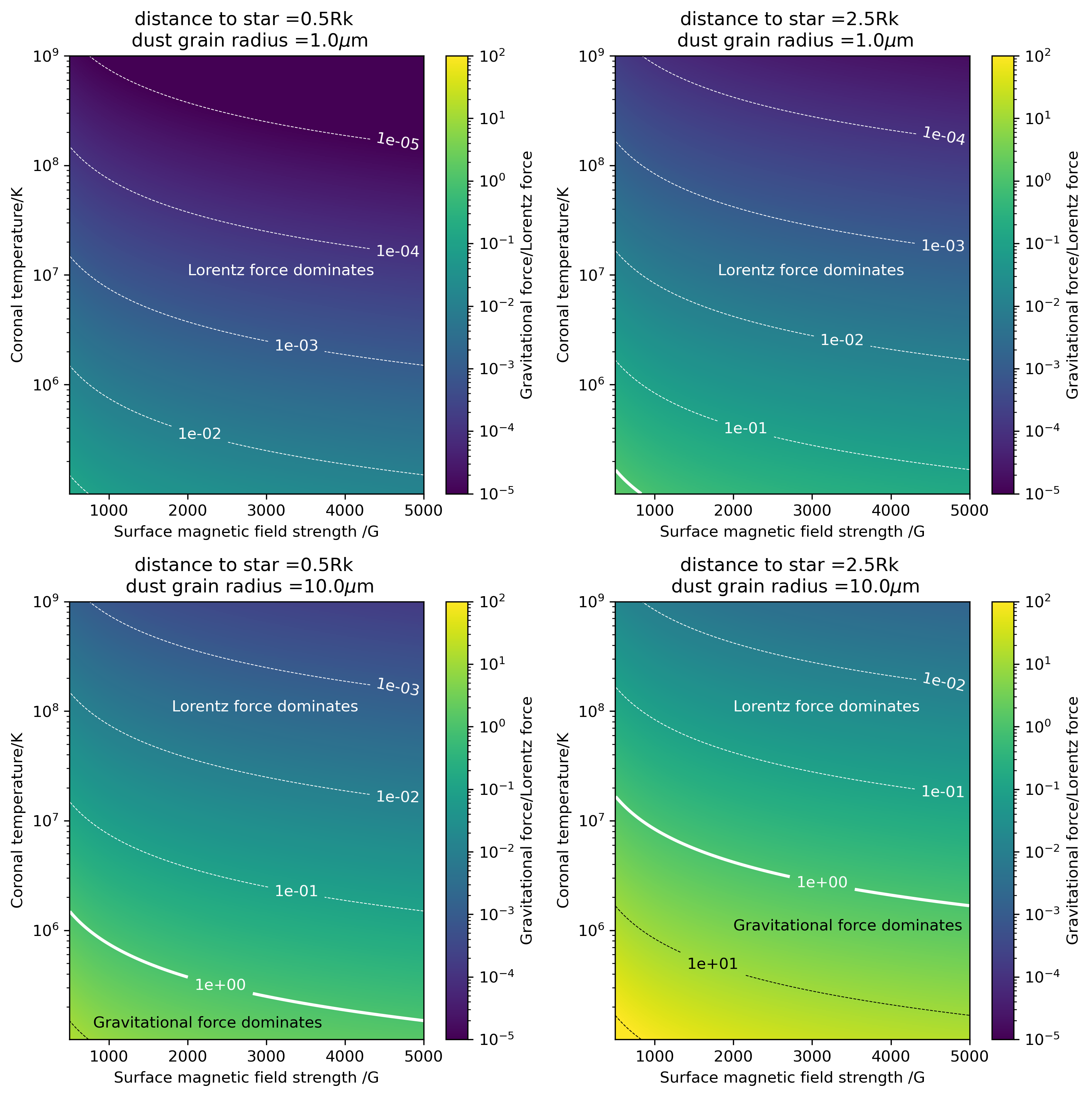}
    \caption{Dust grain support as a function of surface magnetic field strength and coronal temperature, for a variety of grain sizes, $r_d$ and distances from the star, $a$. The distance from the star down each column of the subplot is constant and the grain size across each row of the subplot is constant.}
    \label{fig:support-BT}
\end{figure*}

This indicates dust particles can be supported in the stellar magnetosphere against gravity. The strength of gravitational force relative to the Lorentz force as a function of grain size, distance to the star, surface magnetic field strength and coronal temperature is shown in Figures \ref{fig:support-ar} and \ref{fig:support-BT}. The Lorentz force is larger than the gravitational force except for the largest dust particles furthest from the star. The strength of the Lorentz force in comparison to gravity increases at higher coronal temperatures and surface magnetic field strengths, which increases the maximum grain size that can be supported in the field against gravity.

\begin{figure*}
    \centering
    \includegraphics[width=1\textwidth]{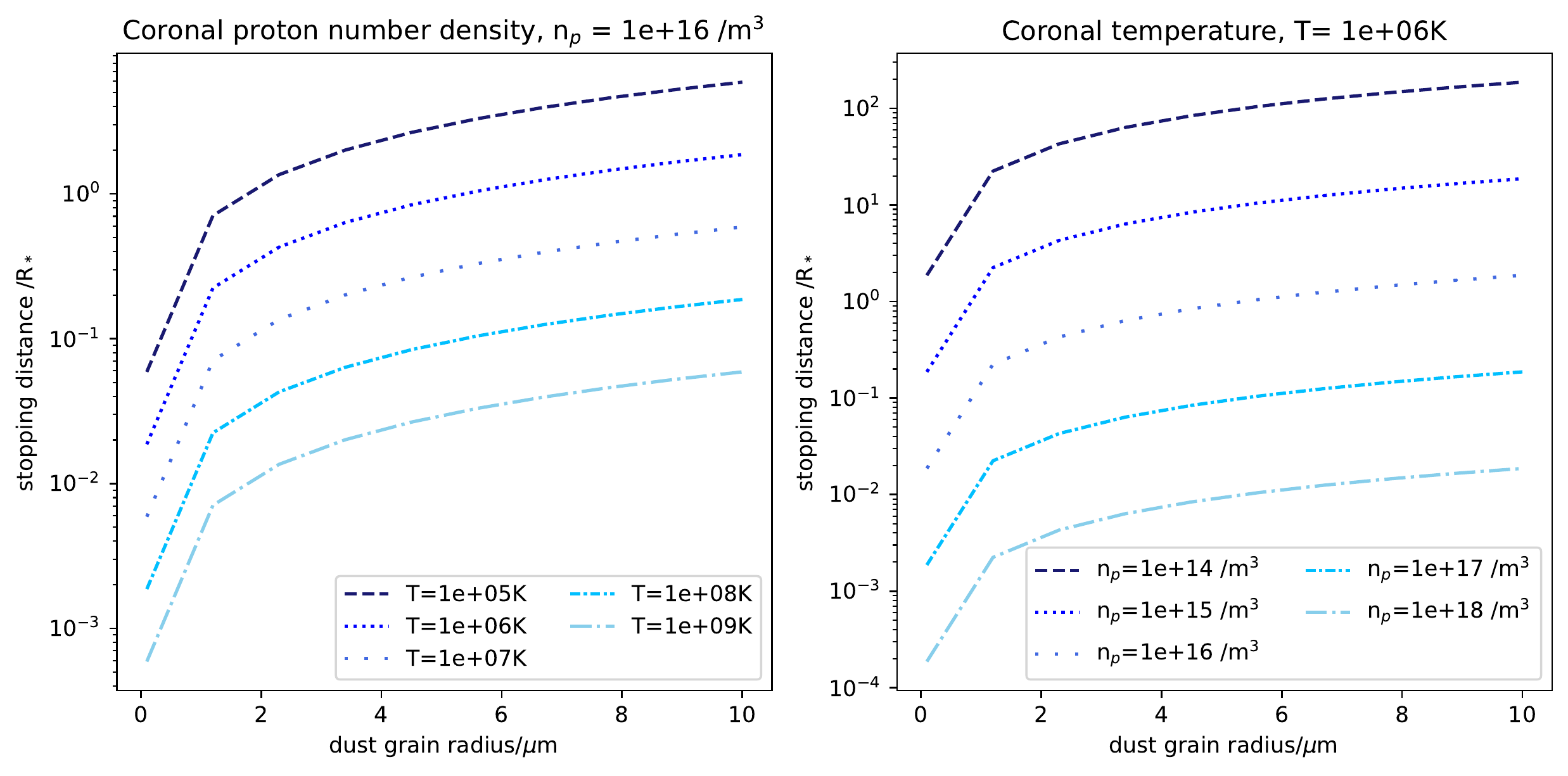}
    \caption{Stopping distance of dust grain as a function of grain size based on Equation \ref{eq:stopping_distance}. In the left panel the number density is fixed at $n_p=10^{16}\,m^{-3}$ and the coronal temperature varies. In the right panel, the temperature is fixed at $T=1\times10^6\,K$ and the coronal number density varies. All other constants have the values in Table \ref{tab:typical_values}}
    \label{fig:stop-dist}
\end{figure*}

Using Equation \ref{eq:stopping_distance}, the approximate stopping distance of a dust grain is \begin{equation}
    d_{\rm s} =0.08\frac{\left(\frac{f}{1}\right)\left(\frac{r_{\rm d}}{1\mu \rm m}\right)\left(\frac{\rho_{\rm d}}{3000\,\rm kg\,\rm m^{-3}}\right)\left(\frac{M_\star}{0.28\,M_\odot}\right)^{\frac{1}{3}}\left(\frac{\Omega_*}{1.63\times10^{-4}\,\rm s^{-1}}\right)^{\frac{1}{3}}}{\left(\frac{n_{\rm p}}{10^{16}\rm m^{-3}}\right)\left(\frac{T}{5.7\, \rm M \rm K}\right)^{\frac{1}{2}}\left(\frac{R_\star}{0.34\,R_\odot}\right)}R_\star.
\end{equation}
The short distance in comparison to the difference between co-rotation radius and cloud distance for the large co-rotation radius in Figure \ref{figure:distances} may mean that only a proportion of the dust grains make it to the stable point, but the distance is not so small that no dust grains could reach the stable points. Additionally, larger dust grains or smaller coronal number densities or lower temperatures could lead to larger stopping distances, as shown in Figure \ref{fig:stop-dist}. A full simulation of the dust motion on the field lines in the corona is needed to improve this estimate of stopping distances and constrain the fraction of dust that settles in a stable point.

The diffusion timescale of the dust trapped at a stable point is \begin{equation}\begin{split}
\tau_{\rm diff}&=2.2\times10^8\left(\frac{r_{\rm cloud}}{0.1\,R_\star}\frac{R_\star}{0.28\,R_\odot}\right)^2\left(\frac{r_{\rm d}}{1\mu \rm m}\right)^2\left(\frac{n_{\rm p}}{10^{16}\, \rm m^{-3}}\right)\\&\left(\frac{T}{5.7\, \rm MK}\right)^{-\frac{1}{2}} \rm yr.
\end{split}\end{equation} This timescale is $10^9$ times larger than the length of observations and indicates once dust is at a stable point it can remain there unless the gas to which it is coupled moves, the dust sublimates (which has short enough timescales for certain compositions) or the field structure changes (see Section \ref{discussion}). Observations of M dwarfs have shown their magnetic field structures evolve only slowly, changing little over timescales of months \citep{vida_investigating_2016}, so these dust clouds could remain fixed for long enough to produce the stable signals observed, contrary to suggestions in \citet{gunther_complex_2022}.

\subsection{Dust-cloud masses and number densities}\label{results:masses} 
The results in Section \ref{results:dust_confinement} combined with the sublimation timescales from \citet{zhan_complex_2019} suggest it is possible to trap dust at stable points in the magnetosphere for long enough to produce the observed absorption. However, the dust masses required must also be physically realistic and the model must reproduce the observed light curves. 

 A typical dust number density in the cloud can be calculated using Equation with 1$\mu m$ dust grains which have $Q_{\rm ext}=3$ and dust-cloud radii from observations in the range $0.06-0.22\,R_\star$ (see Table \ref{table:cloudparams}). \ref{eq:number_densities}, \begin{equation}
    n_{\rm d}=2200\left(\frac{\tau}{1}\right)\left(\frac{Q_{\rm ext}}{3}\right)^{-1}\left(\frac{r_{\rm d}}{1\mu \rm m}\right)^{-2}\left(\frac{s}{0.1\,R_\star}\frac{R_\star}{0.34\,R_\odot}\right)^{-1}\rm m^{-3}.
\end{equation} 
Number densities and values of $Q_{\rm ext}$ are given in Table \ref{tab:n_d}. For a given dust cloud this is a minimum number density as it assumes the longest path length through the cloud and an optical depth of one. This gives a dust to gas number ratio of order $10^{-13}$ relative to the corona. This is consistent with a model of dust particles entrained in coronal plasma and neglecting collisions between dust particles. The dust to gas mass ratio is of order one, so the dust mass is a significant fraction of the mass at a stable point. For a spherical dust cloud with spherical dust grains the cloud mass is \begin{equation}\begin{split}
    m_{\rm cloud}&=1.6\times10^{12}\left(\frac{\tau}{1}\right)\left(\frac{Q_{\rm ext}}{3}\right)^{-1}\left(\frac{r_{\rm cloud}}{0.1\,R_\star}\frac{R_\star}{0.34\,R_\odot}\right)^{2}\\&\left(\frac{r_{\rm d}}{1\mu \rm m}\right)\left(\frac{\rho_{\rm d}}{3000\,\rm kg\, \rm m^{-3}}\right) \rm kg.
\end{split}\end{equation} This is two orders of magnitude smaller than masses of a small rubble-pile asteroid \citep{veras_formation_2014} and many orders of magnitude smaller than typical Vesta (main-belt, rocky) and Eris (Kuiper-belt, rocky) asteroids in the Solar System \citep{2020MNRAS.493..698M}. It is also two orders of magnitude smaller than the mass predicted for cold gas clouds (known as ``slingshot prominences'') trapped at stable points and at least one order of magnitude less than the observed gas mass in a single gas ejection event from V374 Peg \citep{vida_investigating_2016,dangelo_prominence_2018}

\begin{table}
    \centering
    \begin{tabular}{|c|c|c|c|}
    \\\hline
      $r_{\rm d}/\;\mu \rm m$   & $Q_{\rm ext}$ & $n_{\rm d}/\; \rm m^{-3}$ &$m_{\rm cloud}/\; \rm kg$ \\ \hline
        0.1 & 0.2 & $3.4\times10^6$ & $1.6\times10^{9}$
        \\0.5 &2 &$1.3\times10^4$ & $2.0 \times10^{11}$
        \\ 1 & 3& 2200 & $1.6\times10^{12}$
        \\10 & 2& 34 & $1.6\times10^{15}$
        \\\hline
    \end{tabular}
    \caption{$Q_{\rm ext}$ and number densities of dust grains for an optical depth of one for a variety of grain sizes assuming an optical depth of one. }
    \label{tab:n_d}
\end{table}
\subsection{Theoretical light curves}
\label{sec:V374LC}

\begin{figure}
    \centering
    \includegraphics[width=1\columnwidth]{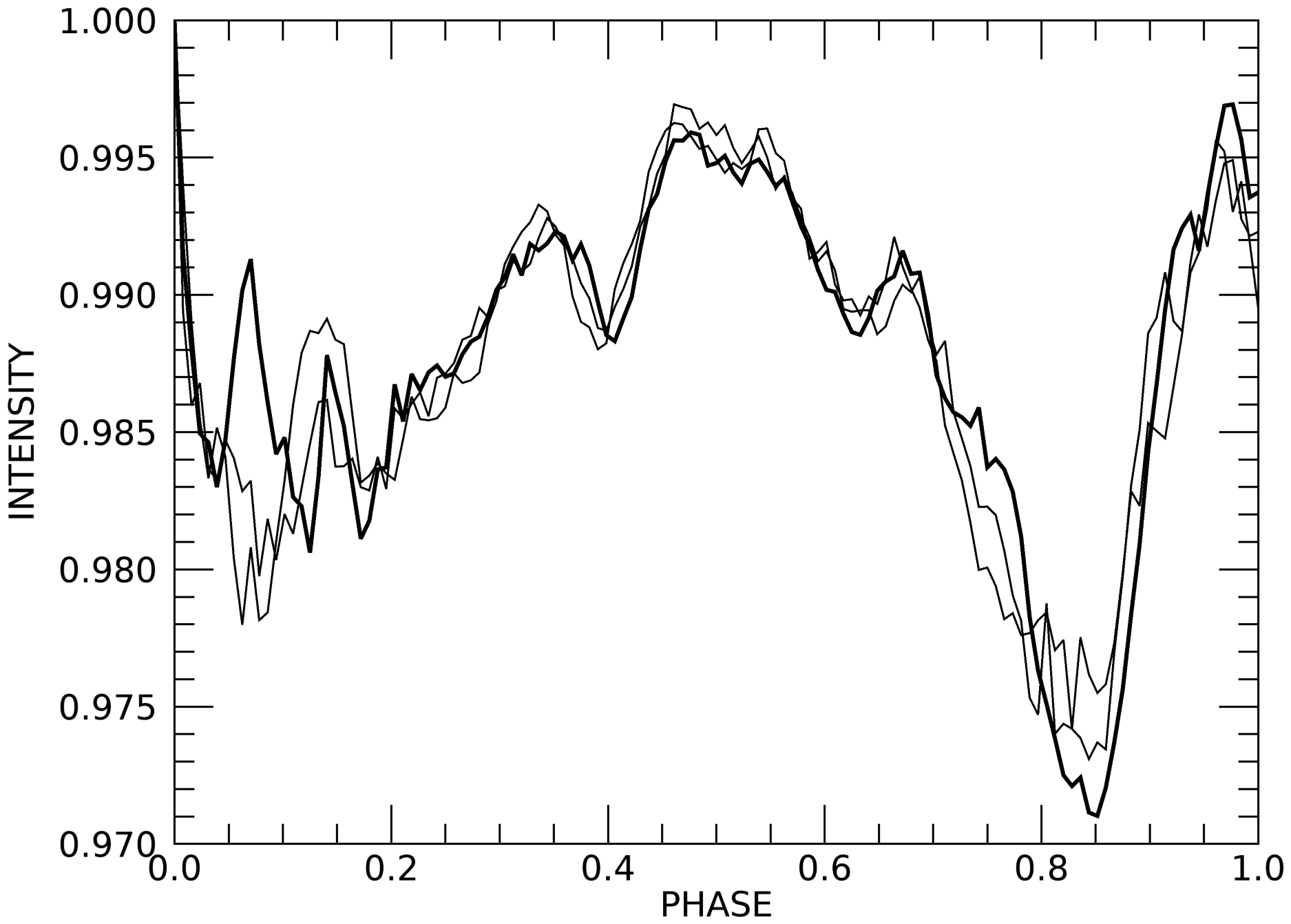}
    \caption{Model light curves for V374Peg created by filling a fraction of the stable points with dust particles of size 1$\mu$m. Three different random samples of stable points are shown. The narrowest features have a width of about 0.05 in phase. The corresponding radial acceleration under-estimates the distance of the cloud from the rotation axis by about 40 percent, as illustrated by the green arrow in Fig.\ref{figure:distances}.}
    \label{fig:v374peg_light_curves}
\end{figure}

Using this number density and corresponding gas to dust ratio, the magnetic map of V374 Peg was used to simulate the light curves. Since not all potential stable points will be filled at any one time, a fraction (50$\%$) of all the stable points was selected at random to support the dust clouds. The locations of all possible stable points and the field structures that support them are shown in Figure \ref{fig:stable_points} and three sample realisations of the resulting light curves for 1$\mu$m dust grains are shown in Fig. \ref{fig:v374peg_light_curves}. For each random selection of stable points, a similar pattern of dips is found. These show depths of 1-3$\%$, consistent with observations.

\begin{figure}
    \centering
    \includegraphics[width=1\columnwidth]{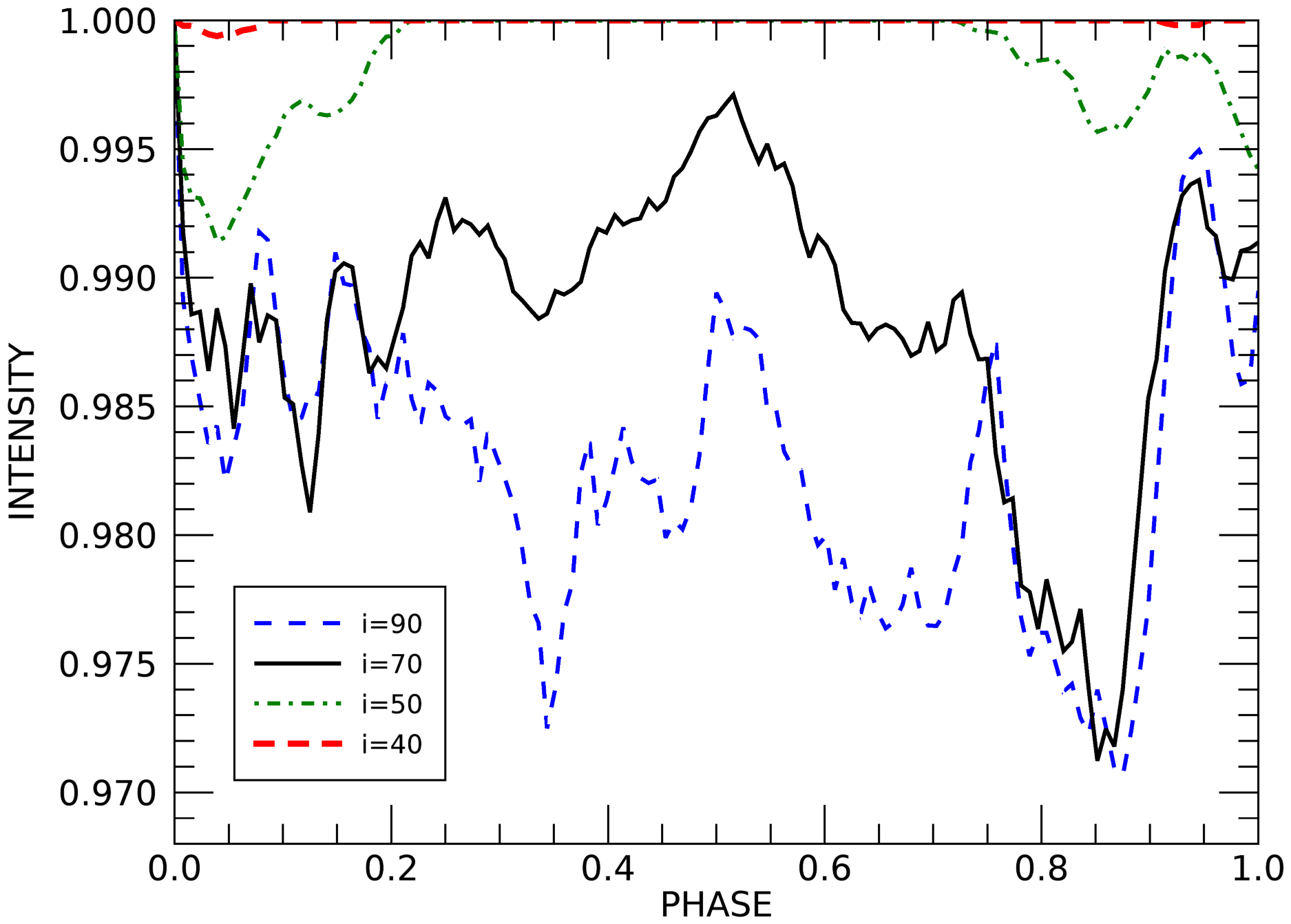}
    \caption{Model light curves for V374Peg created by filling a fraction of the stable points with dust particles of size 1$\mu$m. Four different inclinations of the rotation axis to the line of sight are shown (including $i=70$, the value used in Fig.\ref{fig:v374peg_light_curves}).}
    \label{fig:v374peg_inclinations}
\end{figure}

The inclination of the stellar rotation axis to the line of sight will of course affect which dust clouds transit in front of the star. For a star with a mainly aligned dipolar field, the stable points will lie within a torus that is located around and beyond the co-rotation radius, as shown in Fig. \ref{fig:stable_points}. Reducing the inclination to zero ensures that no dust-clouds will transit (unless they lie over the rotation pole). We therefore expect that the scallop-shell phenomenon will predominantly be found in stars whose rotation axes have large inclinations. We illustrate this in Fig \ref{fig:v374peg_inclinations} which shows light-curves calculated for a range of inclinations, including the value of $i=70^\circ$ \citep{2006Sci...311..633D} used in Fig. \ref{fig:v374peg_light_curves}. For inclinations below $40^\circ$ no dips can be found.\footnote{\citet{2014AJ....147..146K} published projected equatorial rotation speeds for five of the stars in the sample of \citet{zhan_complex_2019} which belong to the Tuc-Hor moving group. Comparison of these $v\sin i$ values with the stellar radii and rotation periods in Table~\ref{table:cloudparams} suggests that all five have rotation axes inclined at less than 40 degrees to the line of sight. One of them, TIC 206544316, was observed independently by \citet{zhan_complex_2019} with spectral resolving power R=24000 on the ANU 2.4-m telescope, yielding vsini=77 km s$^{-1}$. This is very different from the 46.8 km s$^{-1}$  reported by \citet{2014AJ....147..146K} for the same star. The single measurement by \citet{zhan_complex_2019} indicates an inclination close to 90 degrees for TIC 206544316. The reason for the discrepancy is unclear, but we note that the probability of a randomly-chosen sample of 5 stars all having inclinations less than 40 degrees is $7.0\times 10^{-4}$. It would be interesting to re-determine $v\sin i$ for the entire sample.}

\section{Discussion} \label{discussion}%everybody - currently just bullet pointing ideas here, probably best to write properly once have written other sections
\subsection{Wider implications}\label{discussion:implications}
Section \ref{results} demonstrates that dust can be trapped at the co-rotation radius and the resulting dust clouds can produce light curve dips similar to those observed for the scallop-shell stars. The scallop-shell phenomenon is only observed in a small subset of all rapidly-rotating stars with spectral types later than mid-M. This suggests that either the scallop-shell phenomenon is sporadic, or that it is seen only in those stars with favourable viewing geometries.

Many such stars generate strong dipole-dominated fields, which would trap dust close to the equatorial plane. Clouds confined near the stellar equatorial plane will only transit the disc if the inclination of the stellar rotation axis is close to 90 degrees. At low inclinations the clouds may be present but unobservable.

 The probability of a dust cloud at the co-rotation radius crossing the line of sight and causing a dip is \begin{equation}
    P_{\rm transit}=\frac{R_\star+r_{\rm cloud}}{R_K}.
\end{equation} For the 10 scallop-shells discovered by \citet{zhan_complex_2019}, typically $r_{\rm cloud}\simeq 0.1 R_\star$ and $5 R_\star \lesssim R_K \lesssim R_\star$ Hence $P_{\rm transit}$ ranges from 0.22 for $R_K = 5 R_\star$ to 0.055 for $R_K = 20 R_\star$ 
%\begin{equation}
%    P_{\rm transit}=0.26\frac{R_\star+\left(\frac{r_{\rm cloud}}{0.1R_\star}\right)}{\left(\frac{R_K}{4.5R_\star}\right)}.
%\end{equation} 
Therefore, if there was no limit on dust supply, within the subset of young ($<50\, \rm Myr$) and rapidly rotating ($P<1\,\rm day$) M dwarfs, approximately 10\%  would be expected to be scallop-shells. Indeed, \cite{stauffer_more_2018} and \cite{gunther_complex_2022} found 9\% and 6\% of young, rapid rotators were scallop-shells respectively. Nonetheless, \cite{gunther_complex_2022} report that even in stars found to exhibit scallop-shell behaviour, the phenomenon is not always present. This suggests the injection of dust is sporadic, and that the clouds have a finite lifetime. 

Some very low-mass M dwarfs have, however, been shown to generate weaker and multipolar magnetic fields \citep{morin_large-scale_2010}. This may also affect the spatial distribution of stable points where dust can be trapped and hence the transit probability.

 These light curves have only been observed in young stars: 5-10\,Myr for those observed by \citet{stauffer_more_2018} and 45\,Myr for those observed by \citet{zhan_complex_2019} and \citet{palumbo_evidence_2022}. The appearance of this phenomenon in young stars is likely to be because there is more dust and scattering of planetesimals in young stellar systems \citep{williams_protoplanetary_2011}. These stars are also rapid rotators with periods less than one day. Rapid rotation is required to ensure that the co-rotation radius (where most stable points are located) lies within the star's corona. Rapid rotation also ensures that the field is strong enough for the Lorentz force to oppose gravity.

One possible constraint on dust availability is if dust delivery is sporadic. The lack of IR excess in these systems \citep{stauffer_more_2018} indicates the absence of a dusty accretion disc around these stars and so in our model dust moving towards the star is likely to originate from tidally disrupted planetesimals. This is consistent with the dust masses obtained in Section \ref{results:masses}, which are all orders of magnitude lower than asteroid types observed in the Solar System. This supports a model where a small fraction of the planetesimal material reaches the stable point due to the size distribution of planetesimal fragments and the differing effects of gravity and drag on dust grains of different sizes (see Section \ref{results:dust_confinement}). If the rate of tidal disruption of planetesimals around the star is slower than the rate at which dust clouds are destroyed (see Section \ref{discussion:disappearance}) then stars which have not recently had a tidally disrupted planetesimal will not exhibit this phenomenon. 
Tidal disruption of planetesimals around white dwarfs has been modelled extensively to explain white dwarf pollution e.g. \citet{jura_tidally_2003, jura_pollution_2008, veras_formation_2014,malamud_tidal_2020}. These models could be applied to M dwarfs to investigate origins of the dust and determine if tidal disruption is a plausible origin. Resulting theoretical predictions could be tested using  observations of scallop-shell light curves. The frequency of occurrence of scallop shell features could constrain the frequency of planetesimal disruption within the system whilst the timescale for appearance of these features in the light curve (e.g.  between Sectors 10i and 10ii in Figure 9 of \citet{gunther_complex_2022}) could be used to test predicted timescales between disruption and dust accretion. 

Dust trapped in magnetic stable points in these stars provides a new way to map the large-scale loop structures in these stars' coronae. Rather than needing to observe in H$\alpha$ which is costly and can be done for limited time, dust can be observed in the optical with instruments, such as TESS for a long period of time (for example  TIC 177309964 in \citet{gunther_complex_2022} was observed for one year). This could enable high cadence observations of the evolution of M dwarf magnetic fields with these dust clouds acting as a tracer. Magnetic field structures deduced from optical observations could then be compared with those from Zeeman-Doppler imaging.

\subsection{Disappearance of features} \label{discussion:disappearance}

A mechanism for removal of dust clouds is required to explain the lower than expected occurrence rates (Section \ref{discussion:implications}) and the sudden disappearance of the narrow features in the light curve - as  demonstrated between Sector 11ii and Sector 12i in Figure 9 of \citet{gunther_complex_2022}. The sublimation timescales of dust grains are too long to explain this (see \ref{sub-time}).
The magnetically confined, co-rotating dust clouds hypothesis provides a potential process. If the dust is trapped in the magnetic field, the disappearance of the dust signature implies a change in the magnetic field structure. The surface magnetic fields of M dwarfs have been observed to evolve only slowly over timescales of months \citep{2008MNRAS.384...77M} so this change is unlikely to be due to a sudden change in coronal field configuration and disappearance of a stable point. However, this could be explained by the formation (and subsequent ejection) of a ``slingshot prominence'' (a condensation of coronal plasma) at a magnetic stable point. These condensations are observed as travelling absorption features in H$\alpha$ \citep{collier_cameron_fast_1989}. The radial acceleration of these features shows that they typically originate from material at or beyond the co-rotation radius. At these radial distances, they must be supported against centrifugal ejection, most plausibly by the magnetic field.  

The long diffusion timescale in Equation \ref{eq:diffusion} indicates that the dust would be entrained in this condensation. These condensations have observed lifetimes of only around one day. This finite lifetime is a simple consequence of the nature of the gas upflow that forms the prominence \citep{jardine_slingshot_2019}. This upflow is sufficiently hot that it is supersonic by the time it reaches the stable point. As a result, the mass of gas at the stable point will increase until it can no longer be supported by the magnetic field. At this point, any gas (and any entrained dust) will be ejected from the support site. Because the upflow is supersonic, the surface will nonetheless continue to supply mass leading to another ejection. This limit cycle of ejections would ensure the removal of any dust that might have been trapped at the stable point. A similar explanation - centrifugal breakout - has been suggested for disappearance of features in the dipper star TIC 234284556 by \citet{palumbo_evidence_2022}. 

\subsection{Dust particle size}
The size of the dust particles requires further investigation. The upper panels in Figure 10 of \citet{gunther_complex_2022} show weak colour dependence, which could be predominantly due to underlying starspots rather than dust. This dependence and the broadband K2 data suggest the dust grain radii could be anywhere in the visible range or slightly larger. 1$\mu m$ dust grains were suitable for reproducing the signal, as shown in Figure \ref{fig:v374peg_light_curves}. Equation \ref{eq:gravity_lorentz_ratio} indicates an upper limit on particle size of around 5$\mu m$ and Equation \ref{eq:stopping_distance}  demonstrates that larger particles have larger stopping distances and more likely to reach a stable point. Therefore, longer wavelength IR observations are required to confirm the particle size distribution within the range 0.5-5$\mu m$ using methods similar to \citet{croll_multiwavelength_2014}.

\subsection{Further Work} 
The possible coexistence of dust clouds and prominences could be investigated through time-resolved   H$\alpha$ spectroscopy of these stars, with simultaneous photometry. Such a coexistence would depend on rate of prominence formation and dust delivery. Stellar coronae have many stable points, so for frequent dust delivery both scallop-shell light curves and prominences should be observed. This would tell us to what extent the spatial distribution of dust mirrors that of the neutral hydrogen condensations. In addition, if the photometric dips were found always to occur in rotational anti-phase to the H$\alpha$ transients, this would indicate that dips can also arise from bound-free emission being eclipsed as gaseous prominences pass {\em behind} the star, cf. \cite{rebull_rotation_2016}.  

As discussed in Section \ref{discussion:implications}, theoretical work on the delivery of dust to the star is required. As well as models of planetesimal delivery and tidal disruption, it would be instructive to create a magnetohydrodynamic simulation of dust as it enters the corona, including coupling to the field and coronal drag as it travels to a stable point. This would investigate what proportion of dust from a planetesimal will end up in stable points and what the size distribution of this material is.

\section{Conclusions}\label{conclusion} This work has demonstrated that magnetospherically-trapped dust clouds provide a feasible explanation for the high-frequency variability of the scallop-shell stars, seen superposed on the underlying photospheric starspot modulation. Analysis of observations of ten scallop-shells observed by TESS \citep{zhan_complex_2019} has shown that the narrow features appear with the same periodicity as the quasi-sinusoidal starspot modulations and hence are co-rotating. Their transit times are consistent with their being located at or inside the co-rotation radius. We have developed a model for coupling of electrostatically-charged dust particles to the magnetic field, and shown that the existence of magnetic stable points enables trapping of dust around the co-rotation radius. The long sublimation and diffusion timescales of dust at this location agrees with observations of the stability of this signal. Using the magnetic map of V374 Peg, we have generated synthetic realisations of the light curves that would be produced by these co-rotating dust clouds transiting the stellar disc. These have similar features (depth, duration and number of dips) to the observed light curves.

All of the above confirms that co-rotating dust clouds are a possible explanation for scallop-shell stars. For the phenomenon to be observed in a given star requires both a dust delivery mechanism and a favourable viewing geometry. Observations of these stars could provide a test for theoretical models of tidal disruption and planetary dynamics of these systems. The presence of these dust clouds also provides a new method for observing the magnetic stable points of these stars, which could be compared with H$\alpha$ observations and results from Zeeman-Doppler imaging. Further multi-wavelength observations of these stars as well as the development of models of dust delivery processes are important to confirm the validity of this hypothesis and to better understand these stars.

\section*{Acknowledgements}

 The authors would like to thank the Royal Society of Edinburgh for the RSE Cormack Vacation Scholarship which initially enabled this work to be carried out. The authors thank the reviewer for their helpful comments which improved the quality of the manuscript. They would also like to thank Rose Waugh, Doug Lin and Amy Bonsor for helpful discussions. MMJ and ACC acknowledge support from STFC consolidated grant number ST/R000824/1. HS acknowledges funding on NERC studentship NE/S007474/1 and a graduate scholarship from Exeter College, University of Oxford. JFD acknowledges funding from the European Research Council (ERC) under the H2020 research and innovation programme (grant agreement \#740651 NewWorlds). For the purpose of open access, the authors have applied a Creative Commons Attribution (CC BY) licence to any Author Accepted Manuscript version arising.

%%%%%%%%%%%%%%%%%%%%%%%%%%%%%%%%%%%%%%%%%%%%%%%%%%
\section*{Data Availability}

Archival data underpinning the plots is available at polarbase (http://polarbase.irap.omp.eu).



\bibliographystyle{mnras}
\bibliography{Bibliography} % if your bibtex file is called example.bib

%%%%%%%%%%%%%%%%%%%%%%%%%%%%%%%%%%%%%%%%%%%%%%%%%%

%%%%%%%%%%%%%%%%% APPENDICES %%%%%%%%%%%%%%%%%%%%%
\section{Appendix}
\label{app:A}

The solution to Eqn.\ref{eqn:Laplace} may be written as

\begin{equation}
B_r =  \sum^N_{l=1}\sum^l_{m=-l} 
                    B_{lm}P_{lm}(\theta)f_l(r,r_{ss})r^{-(l+2)}e^{im\phi} 
\label{br}
\end{equation}
\begin{equation}
B_\theta   =   -   \sum^N_{l=1}\sum^l_{m=-l} 
            B_{lm}\frac{dP_{lm}(\theta)}{d\theta}g_l(r,r_{ss})r^{-(l+2)}    e^{im\phi}
\label{btheta}
\end{equation}
\begin{equation}
B_\phi   =  - \sum^N_{l=1}\sum^l_{m=-l} 
                        B_{lm}\frac{P_{lm}(\theta)}{\sin\theta} im g_l(r,r_{ss})r^{-(l+2)} e^{im\phi}          
\label{bphi}
\end{equation}
where the functions $ f_l(r,r_{ss})$ and $g_l(r,r_{ss})$  which describe the modification of the field structure by the outflowing hot gas are given by
\begin{equation}
 f_l(r,r_{ss}) = \left[ 
        \frac{l+1+ l(r/r_{ss})^{2l+1}}{l+1+l(1/r_{ss})^{2l+1}}
            \right]
\end{equation}
\begin{equation}
 g_l(r,r_{ss}) =  \left[
       \frac{1 - (r/r_{ss})^{2l+1}}{l+1+l(1/r_{ss})^{2l+1}}
               \right].
\end{equation}

%%%%%%%%%%%%%%%%%%%%%%%%%%%%%%%%%%%%%%%%%%%%%%%%%%

% Don't change these lines
\bsp	% typesetting comment
\label{lastpage}
\end{document}